\documentclass[a4paper,11pt]{article}
\pdfoutput=1
\usepackage{jcappub}
\makeatletter
\let\@fpheader\@empty
\makeatother
\usepackage{booktabs,placeins}
\hypersetup{pdfauthor={Anirudh Pradhan, K. Ghaderi, Sardor Sabirov, Javlon Rayimbaev},pdftitle={Tidal web stability across smoothing scales},pdfkeywords={cosmic web, cosmological perturbation theory, cosmological simulations}}
\newcommand{\EE}{\mathbb{E}}
\newcommand{\PP}{\mathbb{P}}
\newcommand{\tr}{\operatorname{tr}}
\newcommand{\doi}[1]{\href{https://doi.org/#1}{\nolinkurl{doi:#1}}}
\title{Tidal web stability across smoothing scales}

\author[1]{Anirudh~Pradhan,}
\author[2,*]{K.~Ghaderi,\note[*]{Corresponding author.}}
\author[3]{Sardor~Sabirov,}
\author[4,5,6]{and Javlon~Rayimbaev}

\affiliation[1]{Centre for Cosmology, Astrophysics and Space Science (CCASS),\\GLA University, Mathura 281406, U.P., India}
\affiliation[2]{Department of Physics, Mari.C., Islamic Azad University, Marivan, Iran}
\affiliation[3]{Department of General Professional Sciences, Mamun University, Khiva, Uzbekistan}
\affiliation[4]{Harbin Institute of Technology, Harbin 150001, People's Republic of China}
\affiliation[5]{University of Tashkent for Applied Sciences, Str.~Gavhar 1, Tashkent 100149, Uzbekistan}
\affiliation[6]{Institute of Theoretical Physics, National University of Uzbekistan,\\Tashkent 100174, Uzbekistan}

\emailAdd{k.ghaderi@iau.ac.ir}

\abstract{Tidal classifications depend on the smoothing radius even when the volume fractions of voids, sheets, filaments and knots are unchanged. For homogeneous and isotropic Gaussian fields, we derive the leading probability that a site changes class as the radii approach one another and an upper bound at finite separation. We then compare gravitationally evolved maps with Gaussian fields at the same tidal covariance. At zero threshold, sign reversal eliminates the linear response of class disagreement in the growth amplitude. For the reference spectrum, we evaluate the quadratic coefficient relative to a Gaussian field with the ensemble covariance. The coefficient receives contributions from the connected fourth cumulant and the square of the third. The retention contrast between extreme voids and knots at equal abundance has a linear term. In one independent Sahyadri volume at redshift zero, the Gaussian reference overestimates disagreement by 1.09 percentage points at the primary radii and gives excess disagreement at all five thresholds tested.}
\begin{document}
\maketitle

\section{Introduction}
The tidal classification assigns each position to a void, sheet, filament or knot according to the number of eigenvalues of the smoothed gravitational tidal tensor that exceed a threshold \cite{forero}. Other web finders use different fields or geometric criteria \cite{libeskind}. In a homogeneous and isotropic Gaussian density field, the four volume fractions are independent of smoothing radius if the threshold is fixed in units of the smoothed density rms \cite{doro,ayco}. Individual sites can still change class, and the joint distribution of the tensors at the two radii determines how often they do.

Desjacques derived the joint Gaussian shear law at two smoothing radii \cite{des}, and Desjacques and Smith studied shear correlations between distinct positions \cite{ds}. Ay\c{c}oberry et al.\ and Garc\'ia-Bellido used the tidal eigenvalue distribution to study class abundances and information at one radius \cite{ayco,garcia}. Yaryura et al.\ examined the variation of assigned galaxy environments with scale \cite{yaryura} and described those environments with continuous eigenvalue ratios \cite{yaryura_cont}. Related methods identify structures across a range of scales or follow critical points as the smoothing changes \cite{cautun,cadiou}. Here the sites are fixed in Eulerian space and sampled uniformly in volume.

Our Gaussian result is a law for the probability of changing class at nearby radii and a bound at finite separation. Both are determined by the eigenvalue densities at the threshold (section~\ref{sec:stability}). We apply the result to interpret evolved maps against Gaussian and phase controls and measure whether extreme voids and knots retain the same fraction of sites at equal abundance (sections~\ref{sec:evolved} and \ref{sec:shape}). At zero threshold the two observables have different parity under inversion of the initial modes. This leads to responses at different orders, which we calculate in perturbation theory and test at weak amplitude (sections~\ref{sec:weak} and \ref{sec:even}). Finally, we compare two expansion histories at fixed linear growth and repeat the measurements in an independent Sahyadri volume (sections~\ref{sec:history} and \ref{sec:external}).

\section{Scale dependence of the tidal environment}
\label{sec:stability}

\subsection{Joint distribution at two radii}
Consider a centred, homogeneous and isotropic Gaussian density contrast with power spectrum $P(k)$, defined by
$\langle\delta(\boldsymbol k)\delta^*(\boldsymbol k')\rangle
=(2\pi)^3\delta_{\rm D}(\boldsymbol k-\boldsymbol k')P(k)$.
For real isotropic windows $W_R$ with finite, positive variances, define
\begin{equation}
 S_{ab}=\int_0^\infty\frac{k^2\,dk}{2\pi^2}P(k)W_{R_a}(k)W_{R_b}(k),
 \qquad \sigma_a^2=S_{aa},\qquad \rho=\frac{S_{12}}{\sigma_1\sigma_2}.
 \label{eq:rho}
\end{equation}
With $\nabla^2\Phi_R=\delta_R$, define
$X_{a,ij}=\partial_i\partial_j\Phi_{R_a}/\sigma_a$. Its trace is the smoothed density contrast divided by its rms. The angular average of $\hat k_i\hat k_j\hat k_m\hat k_n$ gives
\begin{equation}
 \langle X_{a,ij}X_{b,mn}\rangle
 =\frac{S_{ab}}{15\sigma_a\sigma_b}
 (\delta_{ij}\delta_{mn}+\delta_{im}\delta_{jn}+\delta_{in}\delta_{jm}).
 \label{eq:cov}
\end{equation}
At the same position the joint Gaussian law therefore has the representation
\begin{equation}
 (X_1,X_2)\ \overset{d}{=}\ (A,A\cos\theta+B\sin\theta),\qquad
 \theta=\arccos\rho,
 \label{eq:pair}
\end{equation}
where $A$ and $B$ are independent copies of the normalised tidal tensor. The law is invariant under a common positive rescaling of the density amplitude.

Order the eigenvalues $\lambda_1\geq\lambda_2\geq\lambda_3$ and set
$N_\nu(X)=\#\{j:\lambda_j(X)>\nu\}$. Values $0,1,2,3$ denote a void, sheet, filament and knot, respectively. The common threshold $\nu$ is applied to the normalised tensors at the two radii. In unnormalised units this cut is $\nu\sigma_a$ at radius $R_a$. Holding an unnormalised nonzero threshold fixed generally gives a different normalised cut at each radius. The two conventions coincide at zero threshold. Define
\begin{equation}
 J_{ij}=\PP\{N_\nu(X_1)=i,N_\nu(X_2)=j\},\qquad
 D_\nu(\rho)=1-\sum_{i=0}^3 J_{ii}.
 \label{eq:observable}
\end{equation}
Each entry of $J$ is a joint volume fraction. In the Gaussian pair, the row and column marginals equal the same class fractions, and exchange of the two radii gives $J_{ij}=J_{ji}$. Independent maps have $D_\nu(0)=1-\sum_i\pi_i^2$, and identical maps have $D_\nu(1)=0$.

\subsection{Coherence from class agreement}
At a fixed finite $\nu$, $D_\nu$ decreases strictly on $0\leq\rho\leq1$. Express $A$ in six whitened Gaussian coordinates. Let $f_i$ be the indicator of class $i$ and let $f_{i,m}$ denote its orthogonal projection onto Hermite polynomials of total degree $m$. The Gaussian correlation identity \cite{janson} applied to eq.~\eqref{eq:pair} gives
\begin{equation}
 1-D_\nu(\rho)=\sum_{m=0}^\infty c_m\rho^m,\qquad
 c_m=\sum_{i=0}^3\|f_{i,m}\|_2^2,\qquad
 c_0=\sum_i\pi_i^2,\qquad \sum_{m\geq0}c_m=1 .
 \label{eq:hermite}
\end{equation}
All coefficients are nonnegative. The Gaussian measure gives positive weight to every class, so $c_0<1$ and at least one coefficient of positive degree is nonzero. This proves strict monotonicity, including at the endpoints by continuity. On this interval the population disagreement determines the spectral ratio \(\rho\) in eq.~\eqref{eq:rho}.

The coefficient linear in $\rho$ depends only on the marginal class fractions. In whitened coordinates the trace $t=\tr A$ has unit variance and is independent of the five traceless components. Rotational invariance makes the projection of a class indicator onto every linear traceless component vanish. Giving $t$ a small mean $m$ shifts the effective threshold to $\nu-m/3$. Differentiating the class probability in $m$ gives $\EE[t f_i]=-\pi_i'(\nu)/3$, hence
\begin{equation}
 c_1=\frac{1}{9}\sum_{i=0}^3 [\pi_i'(\nu)]^2,\qquad
 \pi_i'(\nu)=p_{i+1}(\nu)-p_i(\nu),
 \label{eq:weak}
\end{equation}
where $p_1,p_2,p_3$ are the ordered eigenvalue densities and $p_0=p_4=0$. At zero threshold $D_0(0)=0.633959$ and $c_1=0.142453$. For any $0\leq\rho\leq1$, positivity of the remaining Hermite coefficients brackets the disagreement between $D_\nu(0)-c_1\rho-[D_\nu(0)-c_1]\rho^2$ and $D_\nu(0)-c_1\rho$. This bracket requires only the class fractions and their threshold derivatives.

\subsection{The response near equal radii}
For $0\leq\theta\leq\pi/2$ the endpoint disagreement obeys
\begin{align}
 D_\nu(\cos\theta)&= K_\nu\theta+o(\theta)\qquad(\theta\to0),&
 D_\nu(\cos\theta)&\leq K_\nu\theta,\label{eq:law}\\
 K_\nu&=\sqrt{\frac{2}{5\pi}}\sum_{j=1}^3p_j(\nu)
 =\frac{1}{\pi}\left[\sqrt{6}\,e^{-15\nu^2/4}
 +\frac{20\nu^2-1}{3}e^{-5\nu^2/2}\right].\label{eq:closed}
\end{align}
The three terms in the sum for $K_\nu$ are the contributions of the three class boundaries. The closed form follows from the ordered eigenvalue densities of Lee and Shandarin \cite{lee}. In particular $K_0=(\sqrt{6}-1/3)/\pi=0.6735935\ldots$. The central boundary, between sheets and filaments, accounts for $57.88\%$ of the leading class disagreement at $\nu=0$. This percentage uses the boundary weight $\sqrt{2/(5\pi)}\,p_2(0)$ and counts both directions of change.

For a derivation, write $A=H+tI/3$, where $H$ is traceless and $t=\tr A$ is a standard normal variable independent of $H$ and $B$. Conditional on $H,B$, class changes occur in intervals of $t$. The three boundaries for the first tensor lie at $b_j=3[\nu-\lambda_j(H)]$. For the second tensor, at $\theta$ close to zero, they lie at
\begin{equation}
 b_j(\theta)=\frac{3}{\cos\theta}
 \left[\nu-\lambda_j(H\cos\theta+B\sin\theta)\right],
 \qquad b_j'(0)=-3\,v_j^{\mathsf T}Bv_j ,
 \label{eq:boundary}
\end{equation}
where $v_j$ is the unit eigenvector of $H$ associated with $\lambda_j(H)$. The eigenvalues of $H$ are distinct with probability one. For each such realisation the three swept intervals are disjoint at sufficiently small angle. Conditional on $H$, the normal component $v_j^{\mathsf T}Bv_j$ is centred Gaussian with variance $1/5$ by eq.~\eqref{eq:cov}. If $\phi$ is the standard normal density, the marginal at the boundary is $p_j(\nu)=3\EE[\phi(b_j)]$. The eigenvalue perturbation bound controls the sum of interval lengths divided by $\theta$ by a constant times $1+|\nu|+\|H\|+\|B\|$ near zero. Dominated convergence permits the limit to pass through the expectation, giving the first relation in eq.~\eqref{eq:law}.

To bound the disagreement at finite angle, divide the path $A\cos s+B\sin s$ into $m$ equal steps. Different endpoint labels require at least one successive pair to disagree. Rotational invariance in the pair $(A,B)$ makes the probability for each step $D_\nu(\cos(\theta/m))$. The union bound and the small angle limit give eq.~\eqref{eq:law}. This use of a Gaussian rotation is related to bounds on changes of polynomial threshold functions \cite{kane}. The bound concerns the endpoint labels. A return to the initial class contributes zero to this endpoint statistic.

\subsection{Dependence on the smoothing window}
For $P(k)\propto k^n$ with $n>-3$, set $h=\ln(R_2/R_1)$ and $\alpha=(n+3)/2$. A Gaussian window $W_R=\exp[-(kR)^2/2]$ and a sharp Fourier cutoff $W_R=\mathbf{1}_{kR<1}$ give, respectively,
\begin{align}
 \rho_{\rm G}&=(\operatorname{sech}h)^\alpha,&
 D_\nu(\rho_{\rm G})&=K_\nu\sqrt{\alpha}\,|h|+o(|h|),\label{eq:gauss}\\
 \rho_{\rm S}&=e^{-\alpha|h|},&
 D_\nu(\rho_{\rm S})&=K_\nu\sqrt{2\alpha|h|}+o(\sqrt{|h|}).\label{eq:sharp}
\end{align}
The Gaussian window reweights modes smoothly. Moving the sharp cutoff adds or removes a shell of independent modes with variance of order $|h|$, as in excursion set models with independent increments \cite{musso}. For a window differentiable in the spectral norm of eq.~\eqref{eq:rho}, the coefficient of $|h|$ is $K_\nu\|\partial_{\ln R}u_R\|$, where $u_R=W_R/\sigma_R$. This follows from $\|u_{R_2}-u_{R_1}\|^2=2(1-\rho)$.

\begin{figure}[tp]
\centering
\includegraphics[width=\linewidth]{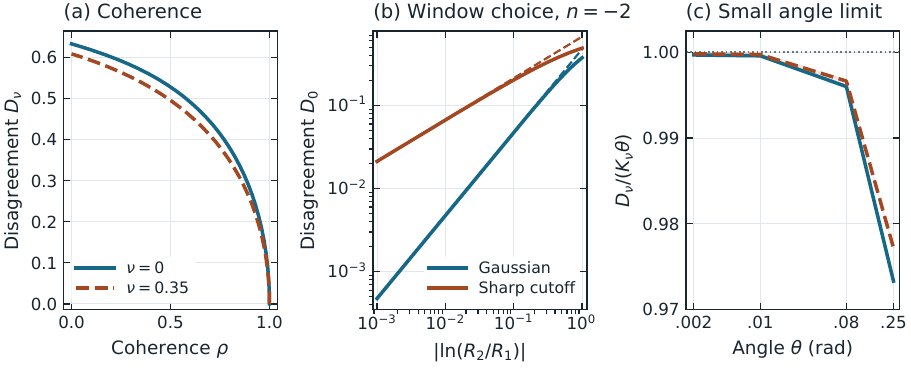}
\caption{Gaussian tidal classification at two smoothing radii. (a) Class disagreement versus the correlation of the two smoothed density fields. (b) Disagreement at zero threshold for Gaussian and sharp Fourier windows at $n=-2$. Solid curves interpolate the pair integrals. Dashed curves show the leading laws in eqs.~\eqref{eq:gauss} and \eqref{eq:sharp}. (c) Independent angular integrations normalised by the leading term $K_\nu\theta$. Colours and line styles follow (a). The bands give one integration standard error.}
\label{fig:stability}
\end{figure}

We compare the coherence dependence at two thresholds in figure~\ref{fig:stability}(a). The two windows are compared at the same fractional change in radius in panel (b). For $n=-2$, $\nu=0$ and $R_2/R_1=1.10$, the fractions are $0.04517\pm0.00008$ and $0.19835\pm0.00031$ for Gaussian and sharp windows, respectively. The class fractions at each radius are identical in both examples. Panel (c) compares the independent angular integrations with the leading law for both thresholds.

\subsection{Numerical evaluation and field checks}
For $-1<\rho<1$, evaluate $J$ by conditioning on the traceless parts $H_a$ of $X_a$ and the half difference $d=(\tr X_2-\tr X_1)/2$. Their common trace variable $c=(\tr X_1+\tr X_2)/2$ is independent of the conditioned variables, with variance $s^2=(1+\rho)/2$. The variable $d$ has variance $(1-\rho)/2$. The increasingly ordered boundaries are
\begin{equation}
 b_{1j}=3[\nu-\lambda_j(H_1)]+d,\qquad
 b_{2j}=3[\nu-\lambda_j(H_2)]-d,\qquad j=1,2,3 .
\end{equation}
Adding $b_{a0}=-\infty$, $b_{a4}=+\infty$ gives the class interval $I_{ai}=(b_{ai},b_{a,i+1})$. If $(L,U)$ are the maximum of the lower endpoints and the minimum of the upper endpoints of $I_{1i},I_{2j}$, the standard normal distribution function $\Phi_{\rm N}$ gives
\begin{equation}
 J_{ij}=\EE\!\left[\max\{0,\Phi_{\rm N}(U/s)-\Phi_{\rm N}(L/s)\}\right].
 \label{eq:integrated}
\end{equation}

We evaluated the remaining expectation in 20 independent batches of 12\,000 pairs. Direct classification of a separate sample gives $0.04498\pm0.00028$ and $0.19778\pm0.00089$ for the two examples above. All quoted sampling errors in this section are one standard error. Independent scrambled integrations at an angle of 0.002 radians reproduce the leading coefficient for both thresholds. Integrating the ordered Doroshkevich density over its two positive eigenvalue gaps reproduces eq.~\eqref{eq:closed} to within $2\times10^{-12}$ at 25 thresholds between $-1.5$ and $1.5$.

We also test the Gaussian prediction in spatially correlated fields (table~\ref{tab:boxes}). We generated periodic realisations of a Gaussian scalar field with $n=0$, removed the constant mode and the Nyquist planes, and applied $k_i k_j/k^2$ to the filtered modes. Zero threshold avoids any estimate of the density rms in assigning labels. The Gaussian smoothing radii are 2.0 and 2.6 grid spacings. In each box we measured the fraction of sites that changed class. Uncertainties use independent boxes as the sampling units. We sum the tensor covariance over the retained lattice modes to obtain the matched prediction. Equation~\eqref{eq:integrated} then integrates the common trace. Cubic symmetry suffices for its independence of the traceless components.

\begin{table}[tp]
\centering\small
\caption{Field realisations and matched lattice predictions for $R_2/R_1=1.3$, $n=0$ and $\nu=0$. The two field columns use disjoint sets of random fields. Parentheses give the standard error in the last digits.}
\label{tab:boxes}
\begin{tabular}{@{}lcccc@{}}
\toprule
Grid & $\rho$ (lattice) & First 20 boxes & Further 80 boxes & Lattice prediction \\
\midrule
$64^3$ & 0.950287 & 0.20662(64) & 0.20501(41) & 0.20482(15) \\
$96^3$ & 0.950251 & 0.20449(42) & 0.20452(22) & 0.20472(18) \\
\bottomrule
\end{tabular}
\end{table}

In the first 20 smaller boxes the mean exceeds the matched prediction by 2.8 combined standard errors. We retained this result and tested 80 further realisations at each size. In the independent sets the means differ from their respective predictions by less than one combined standard error.
For comparison, the isotropic continuum pair gives $D_0=0.20470\pm0.00032$. The Gaussian field tests validate the lattice reference before we apply a realised covariance reference to evolved maps in section~\ref{sec:evolved}.

\section{Gravitational evolution at matched power}
\label{sec:evolved}

Gravitational clustering changes the density power and couples Fourier phases \cite{chiang}. At each epoch we classify the same Eulerian grid at both radii. We compare the measured disagreement with controls that match the power and use rank cuts to compare the retention of extreme voids and knots at equal abundance.

\subsection{Collisionless evolution and initial conditions}
We evolve cold matter in a periodic cube of side $L$ in an Einstein--de Sitter background. With $a$ the scale factor, $H=H_0a^{-3/2}$ and $\boldsymbol p=a^2\dot{\boldsymbol x}$, the particle equations in comoving coordinates are
\begin{equation}
 \frac{d\boldsymbol x}{da}=\frac{\boldsymbol p}{H_0a^{3/2}},\qquad
 \frac{d\boldsymbol p}{da}=\frac{3H_0}{2\sqrt a}\,\mathcal G[\delta](\boldsymbol x),\qquad
 \widehat{\mathcal G[\delta]}(\boldsymbol k)=\frac{i\boldsymbol k}{k^2}\hat\delta(\boldsymbol k),\quad k>0 .
 \label{eq:pm}
\end{equation}
The constant mode has zero force. The initial density extrapolated linearly to $a=1$ is a Gaussian field with power spectrum
\begin{equation}
 P_{\rm L}(k)=A k^{-1}\exp[-(k/6k_0)^2]\,\mathbf1_{0<k\leq 8k_0},\qquad k_0=2\pi/L .
 \label{eq:inputP}
\end{equation}
The symbol $L$ denotes the side of the reference box, including when it sets the comoving scales in a larger box. We set $A$ by requiring an ensemble rms of $0.5$ at $R_1=L/24$ in the reference box at $a=1$. The second radius is $R_2=L/16$. Both windows are Gaussian and $\nu=0$. The cutoff allows the same physical modes to be used in each resolution comparison.

Particles start on a regular lattice displaced at $a_i=1/16$ using the growing solution through second order \cite{crocce}. For $\nabla^2\phi=\delta_{\rm L}$, the initial conditions are
\begin{align}
 \boldsymbol x(\boldsymbol q,a_i)&=\boldsymbol q+a_i\boldsymbol\Psi_1+a_i^2\boldsymbol\Psi_2,
 &\boldsymbol p(\boldsymbol q,a_i)&=H_0a_i^{3/2}(\boldsymbol\Psi_1+2a_i\boldsymbol\Psi_2),\notag\\
 \boldsymbol\Psi_1&=-\nabla\phi,
 &\boldsymbol\Psi_2&=-\frac37\nabla\nabla^{-2}
 \sum_{i<j}(\phi_{,ii}\phi_{,jj}-\phi_{,ij}^2).
 \label{eq:ic}
\end{align}
We pair each linear seed with its negative \cite{pontzen}. The linear displacement changes sign and the quadratic displacement is common to both members. The initial Fourier amplitudes fluctuate with the Gaussian ensemble.

We evolve the system on a particle mesh using cloud in cell deposition and interpolation, a spectral Poisson gradient, and a symmetric kick, drift and kick update. Related force transfer conventions are discussed in ref.~\cite{feng}. The drift and kick factors are the integrals of the corresponding coefficients in eq.~\eqref{eq:pm}. The two kicks bracket the drift and meet at the arithmetic midpoint in $a$. We divide out the two cloud in cell transfer factors in the force and retain $k\leq 2k_{\rm Ny}/3$, where $k_{\rm Ny}$ is the mesh Nyquist wavenumber. For the density measurement we divide out the single deposition factor and discard the constant mode and the Nyquist planes before smoothing. The main ensemble contains 24 independent seed pairs with $48^3$ particles on a $48^3$ mesh and 24 steps equally spaced in $\ln a$. Each box is volume weighted.

\subsection{Two controls for the measured power}
Write $d_{\boldsymbol k}$ for the Fourier series coefficient of the measured density, normalised so that $\delta(\boldsymbol x)=\sum_{\boldsymbol k}d_{\boldsymbol k}e^{i\boldsymbol k\cdot\boldsymbol x}$. For the Gaussian reference we draw centred complex coefficients of variance $|d_{\boldsymbol k}|^2$ on the independent modes and impose the reality condition. The twelve tidal components at the two radii have covariance
\begin{equation}
 C_{ai,bj}=\sum_{\boldsymbol k}|d_{\boldsymbol k}|^2
 W_{R_a}(k)W_{R_b}(k) Q_i(\hat{\boldsymbol k})Q_j(\hat{\boldsymbol k}),
 \label{eq:matchedcov}
\end{equation}
where $(Q_i)=(\hat k_x^2,\hat k_y^2,\hat k_z^2,\hat k_x\hat k_y,\hat k_x\hat k_z,\hat k_y\hat k_z)$ and the sum includes both signs of each mode. This expression retains the directional power of the individual box. At $\nu=0$, normalising each tensor by its trace rms leaves the classification invariant. We evaluate the Gaussian disagreement $D_{\rm G}$ with 65\,536 joint tensor draws per member and epoch. Four further independent sets of Gaussian Fourier coefficients per seed give spatial controls at the linear and final endpoints. Control draws are shared within an inverted pair.

The second control fixes every $|d_{\boldsymbol k}|$ and draws a uniform phase for each independent mode. This preserves the realised power and the coherence of the two smoothed densities. The Gaussian control matches the powers in expectation, so the two references can differ in finite volume. We use two phase realisations per member, shared across epochs. Our primary comparison is the paired difference $\Delta D=D_{\rm map}-D_{\rm G}$.

\subsection{Global agreement and the two extreme environments}
At $a=1$, $D_{\rm map}=0.19715\pm0.00123$ and $D_{\rm G}=0.22392\pm0.00063$. The fixed amplitude phase maps give $0.22394\pm0.00105$. The paired differences are $-0.02677\pm0.00116$ relative to the Gaussian reference and $-0.02679\pm0.00128$ relative to the phase maps. The spatial Gaussian realisations independently give $0.22516\pm0.00097$. Subtracting the residual for the original linear field from the final Gaussian residual gives $-0.02605\pm0.00065$. The linear subtraction controls for the measured offset between a finite Gaussian field and a reference conditioned on its mode powers. All quoted sampling errors use independent seed pairs.

\begin{figure}[tp]
\centering\includegraphics[width=.97\linewidth]{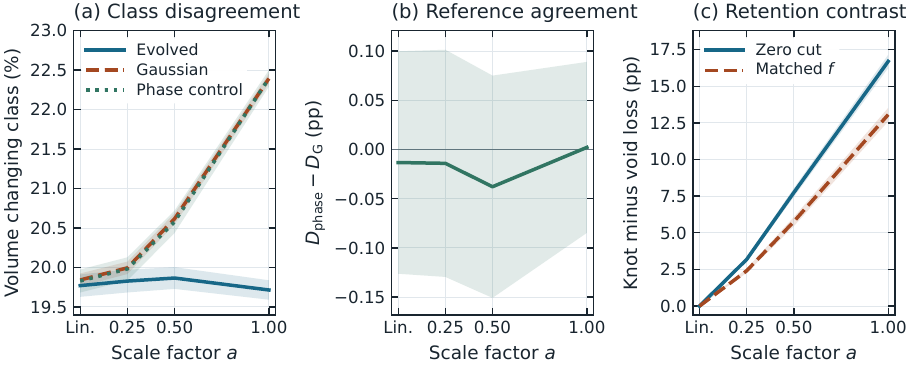}
\caption{Gravitational evolution at fixed radii. (a) Fraction of sites that change class in the evolved map, the Gaussian covariance reference and the fixed amplitude phase control. (b) Phase minus Gaussian disagreement evaluated within each seed pair. (c) Knot minus void loss at zero threshold and at the common abundance $f=0.079676$. Lin. denotes the input classified in the linear limit. Lines join the evaluations. Bands give one standard error over 24 seed pairs.}
\label{fig:evol}
\end{figure}

The discrepancy grows as the coherence of the evolved density falls (figure~\ref{fig:evol}(a)). The mean coherence decreases from $0.95359$ in the linear field to $0.94004$ at $a=1$. In the Gaussian control the disagreement rises from $0.19842$ to $0.22392$. In the evolved maps it stays close to $0.198$ over the same interval. Over this interval the Gaussian reference is more sensitive to the covariance change than the evolved maps are. The phase and Gaussian references agree within their paired errors (figure~\ref{fig:evol}(b)).

Let $\pi_i^{(1)}=\sum_jJ_{ij}$ and $\pi_i^{(2)}=\sum_jJ_{ji}$. The symmetric fraction of class membership lost between the two radii is
\begin{equation}
 \ell_i=1-\frac{2J_{ii}}{\pi_i^{(1)}+\pi_i^{(2)}},\qquad
 C_{\rm KV}=\ell_3-\ell_0 .
 \label{eq:loss}
\end{equation}
The denominator counts membership in both maps and the retained fraction counts the common sites twice. For nonsingular tensors, sign reversal gives $N_0(-X)=3-N_0(X)$. Thus $C_{\rm KV}=0$ in any population whose joint law is invariant under simultaneous sign reversal. This equality holds for the centred Gaussian control even with the anisotropic covariance of an individual box. It also holds in the uniform random phase law because simultaneous sign reversal preserves that law. We evaluate the ratio within each box and average the two members before estimating ensemble means and errors.

\begin{figure}[tp]
\centering\includegraphics[width=\linewidth]{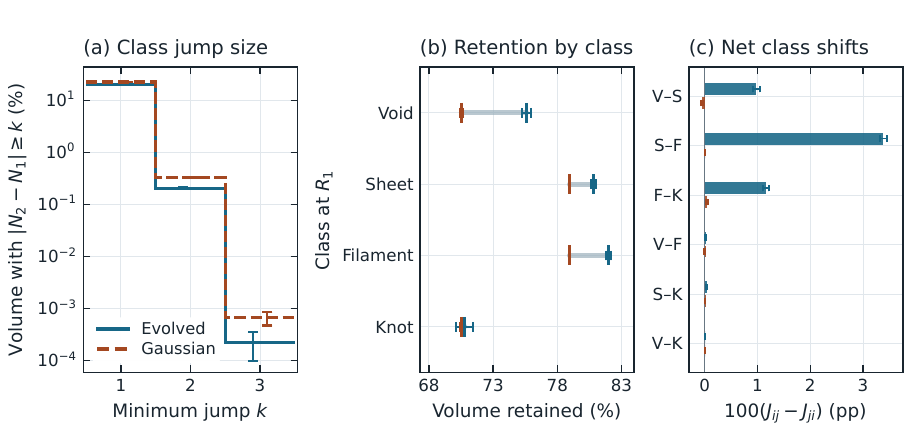}
\caption{Changes in tidal classification between the two radii at $a=1$. (a) Fraction of the volume with $|N_2-N_1|\geq k$, where $N_s$ is the number of positive tidal eigenvalues at $R_s$. (b) Retention at $R_2$ conditional on the class at $R_1$, given by $J_{ii}/\sum_j J_{ij}$. The grey segments connect the two estimates. (c) Excess transitions from class $i$ to class $j$ relative to the reverse transition. A pair labelled S--F, for example, gives the sheet to filament volume minus the filament to sheet volume. V, S, F and K denote void, sheet, filament and knot. The series identification in (a) applies to all panels. Statistics are evaluated for each sign separately, and the two signs are averaged within each seed pair. Error bars show one standard error across the 24 pairs.}
\label{fig:matrix}
\end{figure}

At $a=1$, the evolved maps give $\ell_3=0.3740\pm0.0039$ and $\ell_0=0.2071\pm0.0026$, a contrast of $0.1669\pm0.0045$. The first tail probability in figure~\ref{fig:matrix}(a) is $D$. Class jumps of two or more account for $0.204$ percent of the evolved volume and $0.324$ percent in the Gaussian reference. Panel (b) shows retention conditioned on the class at the inner radius. This differs from the symmetric membership loss in eq.~\eqref{eq:loss}. In panel (c) the largest directional contrast is between sheets and filaments. The fraction classified as a sheet at the inner radius and a filament at the outer radius exceeds the reverse fraction by $3.39\pm0.07$ percentage points. In the full joint matrix, the Gaussian reference underestimates the fraction classified as sheets at both radii by $5.60$ points and overestimates the fraction classified as filaments at both radii by $3.07$ points. The electronic supplement contains all matrix entries.

We now compare the extremes at fixed abundance. At each radius let $K_a(f)$ contain the fraction $f$ of sites with the largest $\lambda_3$, and let $V_a(f)$ contain the fraction $f$ with the smallest $\lambda_1$. Applied separately to these two excursion sets, eq.~\eqref{eq:loss} gives the difference in loss
\begin{equation}
 C_f=\frac{\PP[V_1(f)\cap V_2(f)]-\PP[K_1(f)\cap K_2(f)]}{f}.
 \label{eq:rank}
\end{equation}
On a finite grid we take the same integer count for all four masks and use the corresponding realised fraction. Strictly increasing remappings of the selected eigenvalue at either radius leave the ranks and $C_f$ unchanged.

At $f=0.079676$, the Gaussian abundance of either extreme class at zero threshold, the evolved field gives $C_f=0.1306\pm0.0046$ at $a=1$ (figure~\ref{fig:evol}(c)). The common volume of the two void sets exceeds the common volume of the two knot sets by $fC_f\simeq0.0104$ of the box volume. For $f=0.05$ and $0.10$ the contrasts are $0.1313\pm0.0061$ and $0.1305\pm0.0036$. These rank cuts define two extreme sets. The original four class partition is used for $D$ throughout.

\subsection{Numerical accuracy and physical scope}
We refine the calculation using the same initial modes in comoving coordinates (table~\ref{tab:evolutionchecks}). For the six seeds shared by the $72^3$ and $96^3$ calculations, this further mesh refinement changes $\Delta D$ by $-0.030\pm0.004$ percentage points. Doubling the box side at fixed physical radii, force resolution and input spectral amplitude adds long wavelength modes. Eight independent seed pairs in the larger volume give $\Delta D=-0.02845\pm0.00092$ and $C_{\rm KV}=0.1756\pm0.0030$.

\begin{table}[tp]
\centering\small
\caption{Matched numerical checks at $a=1$. Entries give the change from the reference setting in percentage points with one standard error on that change. Grid refinement increases particle and force mesh resolution together. The earlier start comparison holds the spacing in $\ln a$ fixed.}
\label{tab:evolutionchecks}
\begin{tabular}{@{}lrrr@{}}\toprule
Change & Seed pairs & Change in $\Delta D$ & Change in $C_{\rm KV}$ \\ \midrule
$48^3$ to $72^3$ & 12 & $-0.039\pm0.009$ & $+0.241\pm0.052$\\
$48^3$ to $96^3$ & 6 & $-0.075\pm0.017$ & $+0.375\pm0.059$\\
24 to 48 steps at $48^3$ & 8 & $-0.020\pm0.002$ & $+0.061\pm0.013$\\
$a_i=1/16$ to $1/32$ at $48^3$ & 8 & $+0.012\pm0.003$ & $-0.096\pm0.020$\\
\bottomrule\end{tabular}
\end{table}

We check eq.~\eqref{eq:pm} with a plane wave. The linear input is $0.30\cos(2\pi q/L)$. Before shell crossing, the exact displacement along the wavevector is $-0.30\,aL\sin(2\pi q/L)/(2\pi)$ \cite{taruya}. At $a=1$ and 24 steps, the rms displacement errors relative to the analytic rms are $3.12\%$, $2.04\%$ and $1.27\%$ on $32^3$, $48^3$ and $72^3$ meshes. The corresponding relative errors in the fundamental amplitude are $0.151\%$, $0.127\%$ and $0.119\%$. The tests in table~\ref{tab:evolutionchecks} apply directly to the classification statistics.

The deficit in class changes and the larger overlap of extreme voids persist under these refinements. The next comparison varies the input spectrum and the separation of the smoothing radii.

\section{Spectral shape at fixed radii and fixed coherence}
\label{sec:shape}

Changing the initial spectrum changes the evolved power \cite{smith} and the coherence between any two fixed smoothing windows. We first vary the input spectrum at common comoving radii. We then adjust the radii to match the density rms and coherence, to test whether those two scalars account for the spectral dependence of class disagreement.

\subsection{Fixing the variance and coherence of each realisation}
We replace eq.~\eqref{eq:inputP} by the family
\begin{equation}
 P_n(k)=A_n(k/k_0)^n \exp[-(k/6k_0)^2]\,\mathbf1_{0<k\leq 8k_0},
 \qquad n\in\{-2,-1,0\}.
 \label{eq:tilted}
\end{equation}
For each tilt, $A_n$ sets the ensemble linear rms to $0.50$ at $L/24$ on the reference lattice. The cutoff and the reference wavenumber $k_0=2\pi/L$ are held fixed in comoving units. We use 32 independent Gaussian realisations, with the same standardised Fourier coefficients for all three spectra within each realisation. We evolve each realisation and its sign inverse with the dynamics of section~\ref{sec:evolved}. All results below concern $a=1$ and zero threshold unless a linear control is identified.

For Gaussian smoothing set $t=R^2$ and write
\begin{equation}
 S(t)=\sum_{\boldsymbol k}|d_{\boldsymbol k}|^2 e^{-t k^2},
 \qquad \mu(t)=-\frac{d}{dt}\log S(t).
 \label{eq:heat}
\end{equation}
Thus $\sigma_R^2=S(t)$, $\mu$ has dimensions of inverse length squared, and $\mu'(t)=-\mathrm{Var}_t(k^2)$ under weights proportional to $|d_{\boldsymbol k}|^2e^{-t k^2}$. For $t_2>t_1$ the coherence obeys
\begin{equation}
 \rho=\frac{S((t_1+t_2)/2)}{\sqrt{S(t_1)S(t_2)}},\qquad
 \frac{\partial \log\rho}{\partial t_2}
   =\frac{\mu(t_2)-\mu((t_1+t_2)/2)}{2}.
 \label{eq:monorho}
\end{equation}
The variance decreases strictly with radius when the spectrum contains power at positive wavenumber. The coherence decreases strictly from unity if the power occupies at least two distinct wavenumber magnitudes. Each radius is then uniquely determined provided the target is bracketed. In each realisation we first solve $S(t_1)=0.25$ and then solve eq.~\eqref{eq:monorho} for $\rho=0.940$. Every member of the ensemble reached both targets. The rms at the second radius and the angular distribution of power can still vary across spectra. Within each member the Gaussian control uses the full tensor covariance in eq.~\eqref{eq:matchedcov}, and every phase control uses the same two radii as its parent field.

\subsection{Dependence on the radius separation}
\begin{figure}[tp]
\centering\includegraphics[width=.99\linewidth]{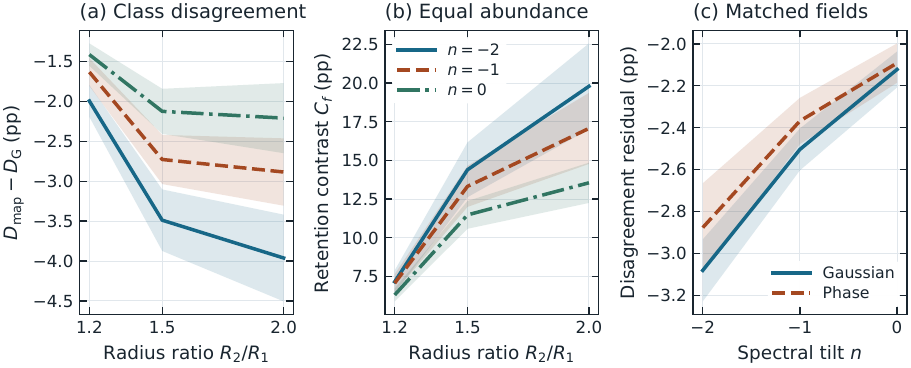}
\caption{Spectral shape and class stability. Panels (a) and (b) show the disagreement residual and equal abundance retention contrast at fixed $R_1=L/24$. The shaded intervals have 95 percent family coverage under a Bonferroni adjusted Student approximation for nine comparisons in each panel. Colours and line styles identify the same spectral tilts in (a) and (b). Panel (c) gives the evolved minus reference disagreement at matched $\sigma_1=0.50$ and $\rho=0.940$ for the Gaussian and phase references. Its bands are one standard error across seed pairs. Lines join evaluated cases.}
\label{fig:shape}
\end{figure}

We compare $\Delta D$ for three fixed values of $R_2/R_1$ at $R_1=L/24$ in figure~\ref{fig:shape}(a). The evolved density changes class less frequently than its Gaussian reference in all nine comparisons. The deficit grows in magnitude from 1.99 to 3.96 percentage points as the radius ratio increases from 1.2 to 2 for $n=-2$. For $n=0$ it grows from 1.41 to 2.21 points. Panel (b) gives the retention contrast at equal abundance, defined in eq.~\eqref{eq:rank}. At the smallest separation its values span 6.30 to 7.16 points. At the largest they span 13.55 to 19.79 points. The nine simultaneous intervals are negative for $\Delta D$ and positive for $C_f$.

At $n=-1$ and $R_2/R_1=1.5$, the new seeds give $\Delta D=-0.02727\pm0.00103$ and $C_f=0.13324\pm0.00446$. The independent estimates in section~\ref{sec:evolved} differ from these by 0.32 and 0.41 combined standard errors, respectively. Errors here and below refer to the dispersion of independent seed pair means divided by the square root of the number of pairs. Both signs, all three tilts and the associated controls stay together when a common seed is resampled.

\subsection{What remains at matched coherence?}
The ensemble means at matched rms and coherence are given in table~\ref{tab:matchedshape}. The disagreement residuals for both references are shown in figure~\ref{fig:shape}(c). In the isotropic Gaussian limit at zero threshold, fixing $\rho$ fixes the joint law of the normalised tensors in eq.~\eqref{eq:pair}. On the finite lattices, $D_{\rm G}$ changes by 0.23 percentage points between the two extreme tilts. The evolved residual changes appreciably more. Taking $n=0$ minus $n=-2$ gives a paired increase of $0.96\pm0.15$ points in $\Delta D$ (95 percent Student interval $[0.66,1.26]$). The disagreement deficit is smaller for the flatter spectrum. The same comparison against the fixed amplitude phase maps gives $0.79\pm0.22$ points. Subtracting each tilt's linear residual before taking the difference gives $0.74\pm0.11$ points against the Gaussian reference. This last comparison allows for the finite sample offset induced by conditioning on the measured powers.

\begin{table}[tp]
\centering\small
\caption{Main ensemble at matched $\sigma_1=0.50$ and $\rho=0.940$. Radii are adjusted for each member. The three columns after $n$ are means over members. Each radius ratio is formed before averaging. The final two columns give means and one paired standard error in percentage points. A negative $\Delta D$ means that the evolved map retains more class labels than the Gaussian control.}
\label{tab:matchedshape}
\begin{tabular}{@{}rrrrrr@{}}\toprule
$n$ & $R_1/L$ & $R_2/R_1$ & $\sigma_2$ & $100\,\Delta D$ & $100\, C_f$\\ \midrule
$-2$ & 0.04715 & 1.605 & 0.351 & $-3.08\pm0.15$ & $14.78\pm0.69$\\
$-1$ & 0.04355 & 1.495 & 0.343 & $-2.50\pm0.10$ & $12.91\pm0.41$\\
$0$  & 0.04117 & 1.428 & 0.326 & $-2.12\pm0.09$ & $10.77\pm0.26$\\ \bottomrule
\end{tabular}
\end{table}

Void regions retain a larger fraction of their volume than knot regions at each tilt. The difference in retention falls by $4.01\pm0.64$ percentage points from $n=-2$ to $n=0$ (signed 95 percent interval $[-5.31,-2.71]$). Equation~\eqref{eq:rank} fixes the marginal abundances, so this difference reflects the joint rank distribution. Randomising the phases gives mean rank contrasts between $-0.10$ and $0.00$ points in the main ensembles, all within one standard error of zero.

Increasing the particle and force resolution from $48^3$ to $72^3$ on eight shared seeds changes $\Delta D$ by at most 0.032 percentage points in absolute mean at any matched tilt. At the finer resolution the difference between $n=0$ and $n=-2$ is $1.17\pm0.24$ points in $\Delta D$ and $-4.78\pm0.88$ points in $C_f$. In a cube of side $2L$ with $96^3$ particles, eight independent pairs per tilt give $0.96\pm0.14$ and $-4.54\pm0.64$ points, respectively. The physical radii needed for the match do depend on volume, especially for $n=-2$, where the mean $R_1$ increases from $0.0472L$ to $0.0593L$. The endpoint contrast retains its sign in both checks.

For these regulated spectra, the smaller evolved disagreement and the larger spatial overlap of extreme voids persist at the same density rms and coherence. Both measured magnitudes depend on the input spectrum. The matched comparison shows that these two summaries do not determine either statistic in the Einstein--de Sitter experiments. Since the match uses different comoving radii at each tilt, the fixed radius comparison is needed to assess spectral dependence at common comoving scales.

\section{Gravitational origin of the difference in tidal retention}
\label{sec:weak}

Reversing both tidal tensors leaves the class change event invariant and exchanges the two extreme environments at zero threshold. This symmetry determines which statistics can respond to the gravitational bispectrum at leading order. We return to $n=-1$ and to radii $L/24$ and $L/16$. We use the population version of eq.~\eqref{eq:rank} with $f=0.079676$, assigning a common marginal fraction at each radius over the ensemble. At zero threshold we also use $A_0$, the signed difference of the joint void and joint knot volume fractions.

\subsection{The third cumulant at two radii}

Let $T$ collect the twelve components of the two tidal tensors formed from $\delta(a)/a$, in the order used in eq.~\eqref{eq:matchedcov}. Dividing by the positive growth amplitude preserves the zero threshold classes and the eigenvalue ranks. Write $L_0L_0^{\mathsf T}$ for the ensemble covariance of the linear tensor on the input lattice, and $Z=L_0^{-1}T$. In the linear limit $Z$ is a standard Gaussian vector. For a growing Gaussian initial density in Einstein--de Sitter space, the first nonlinear term is $\delta(a)/a=\delta_1+a\delta_2+\cdots$. The quadratic kernel of the density is \cite{bern}
\begin{equation}
 F_2(\boldsymbol p,\boldsymbol q)
 =\frac57+\frac{\mu}{2}\left(\frac pq+\frac qp\right)+\frac27\mu^2,
 \qquad \mu=\frac{\boldsymbol p\cdot\boldsymbol q}{pq}.
 \label{eq:treeF}
\end{equation}
For $P_{\boldsymbol p}=\EE|d_{1,\boldsymbol p}|^2$, let $B_i(\boldsymbol p)$ be the linear filter for component $i$ of $Z$. It includes the tidal projection, smoothing and multiplication by $L_0^{-1}$. Set $B_i(\boldsymbol 0)=0$. Wick contraction gives the derivative of the third cumulant at $a=0$,
\begin{equation}
 K_{ijk}=2\!\sum_{\boldsymbol p,\boldsymbol q}P_{\boldsymbol p}P_{\boldsymbol q}
 F_2(\boldsymbol p,\boldsymbol q)
 \big[B_i(\boldsymbol p)B_j(\boldsymbol q)B_k(\boldsymbol p+\boldsymbol q)
       +\text{cyclic permutations of }i,j,k\big].
 \label{eq:treeK}
\end{equation}
The sum runs over both signs of each input mode. Smoothing acts on the output of the quadratic density as well as on the two linear fields in the contraction. The density mean vanishes, and the covariance has zero first derivative because it would involve an odd moment of the Gaussian input.

For a fixed event $E$ of the two tensors, the first variation of its probability is
\begin{equation}
 \left.\frac{d}{da}\PP_a(E)\right|_0=\EE_0[\mathbf1_E H(Z)],\qquad
 H(z)=\frac{K_{ijk}}{6}
 \left(z_i z_j z_k-z_i\delta_{jk}-z_j\delta_{ik}-z_k\delta_{ij}\right),
 \label{eq:score}
\end{equation}
with repeated component indices summed. To see this, write the perturbed vector as $Z+aY$ and condition $Y$ on $Z=z$. Its conditional mean has the form $M_{ijk}(z_jz_k-\delta_{jk})/2$, where $M_{ijk}=\EE_0[Y_i(Z_jZ_k-\delta_{jk})]$. Integrating the first variation of a smooth test function by parts gives eq.~\eqref{eq:score} with $K_{ijk}=M_{ijk}+M_{jik}+M_{kij}$. The density expansion for the finite input lattice holds in $L^1$, so it also applies to the class and rank indicators (appendix~\ref{app:density}). A related calculation predicts the abundances of tidal environments at one smoothing radius \cite{ayco}. Here the mixed cumulants link two radii. Since $H(-z)=-H(z)$ and $N_0(-T)=3-N_0(T)$ almost surely, the derivative of the total disagreement is exactly zero. At a nonzero threshold, inversion sends $N_\nu(-T)$ to $3-N_{-\nu}(T)$. It therefore relates the responses at opposite thresholds and allows a linear correction at either one. The Gaussian reference at ensemble covariance has zero first derivative as well. Reversing the linear modes reverses the first variation of each squared Fourier amplitude, while leaving its zeroth order value unchanged. Thus the disagreement residual in section~\ref{sec:evolved} has no term linear in $a$ about the growing Gaussian limit. The antisymmetric observable $A_0$ can have such a term.

\subsection{Moving thresholds at fixed abundance}

Write $u_s$ for the smallest eigenvalue at radius $s$, and $q_s$ for the threshold with $\PP_0(u_s>q_s)=f$. Let $I_s=\mathbf1_{u_s>q_s}$ and $b_s=\PP_0(I_{3-s}=1\mid u_s=q_s)$. Analogous thresholds for voids use the negative of the largest eigenvalue. For each radius, the knot threshold and the reflected void threshold agree at $a=0$. If $p_s$ is the Gaussian marginal density of $u_s$, differentiating the fixed fraction gives $p_s(q_s)\dot q_s=\EE_0[I_sH]$. A shift of threshold $s$ changes the joint knot probability by $-b_sp_s(q_s)\dot q_s$. Subtracting the corresponding change for the reflected void population yields
\begin{equation}
 c_f=\left.\frac{dC_f}{da}\right|_0
 =-\frac{2}{f}\,\EE_0\big[(I_1I_2-b_1 I_1-b_2 I_2)H\big].
 \label{eq:rankresponse}
\end{equation}
The marginal correction has the form familiar from derivatives of joint rank distributions \cite{segers}. Here their weights multiply the specific gravitational response in eq.~\eqref{eq:score}. For $A_0$ the eigenvalue cuts stay at zero, so $c_A=\dot A_0=-2\EE_0[\mathbf1_{u_1>0,u_2>0}H]$.

We evaluate eq.~\eqref{eq:treeK} on the periodic input lattice. Moments measured in real space provide a check on the Fourier contraction. In the Gaussian integral one independent standard normal coordinate can be chosen to add the same scalar multiple of the identity to both tensors. In whitened coordinates this direction is $L_0^{-1}e/\|L_0^{-1}e\|$, with $e$ representing the identity matrix at each radius. At fixed remaining coordinates, each knot condition is a lower cut on that scalar, and $H$ is a cubic polynomial in it. Integrating its truncated Gaussian moments gives the joint response and the densities at the two boundaries. We integrate the remaining eleven coordinates using 32 independent scrambles of a $2^{17}$ point Sobol sequence, including the sign reflection of each sample. A separate integral in all twelve coordinates tests the boundary calculation.

The joint knot term alone in eq.~\eqref{eq:rankresponse} gives $0.74760\pm0.00075$ per unit $a$. The threshold terms contribute $-0.63756\pm0.00053$. Their sum, evaluated within each integration batch, gives
\begin{equation}
 c_f=0.11004\pm0.00038,\qquad c_A=0.05980\pm0.00006.
 \label{eq:weaknumbers}
\end{equation}
The errors in eq.~\eqref{eq:weaknumbers} describe numerical integration. Fixing the abundance removes about 85 percent of the response inferred by holding the two Gaussian rank thresholds fixed. The remaining linear response favours retention of extreme voids.

\subsection{Test against gravitational evolution}

A test of the perturbative coefficient uses the explicit density $\delta_1+\epsilon\delta_2$ with $\epsilon=0.10$ and 32 fresh pairs of inverted initial fields. Pooling sites before setting the rank thresholds gives $C_f=0.01103\pm0.00069$ and $A_0=0.00586\pm0.00018$. The leading predictions from eq.~\eqref{eq:weaknumbers} are $0.01100$ and $0.00598$. For the dynamical test we evolve the same inputs with eq.~\eqref{eq:pm}. Jackknife errors on $C_f$ use one omitted seed pair per replicate and recompute all four thresholds. At $a=1$, assigning equal abundance separately within each box instead gives $0.13201\pm0.00444$, compared with the pooled value $0.13273\pm0.00467$.

\begin{figure}[tp]
\centering\includegraphics[width=.98\linewidth]{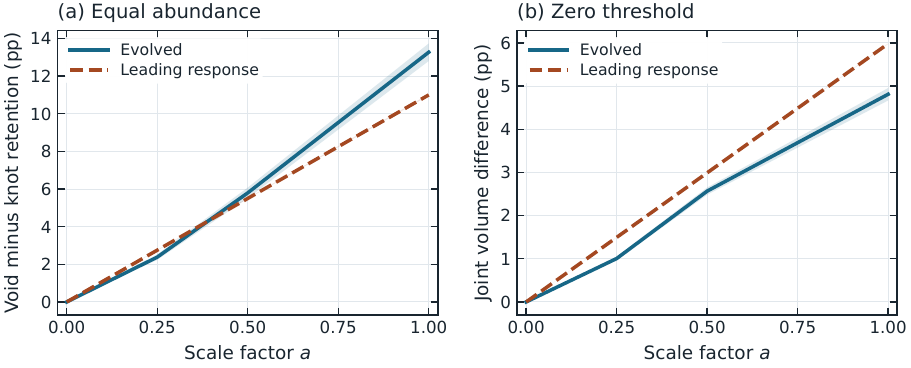}
\caption{Gravitational evolution at the original fixed radii. (a) Difference in retention between extreme voids and knots at equal abundance. (b) Joint void volume fraction minus joint knot volume fraction at zero threshold. Solid lines join the linear limit and the three measured epochs. Dashed lines show the leading predictions from the input power with no fitted amplitude. Bands give one standard error over seed pairs for the maps and over integration batches for the predictions. Both ordinates are in percentage points.}
\label{fig:weak}
\end{figure}

We compare the rank contrast with $a c_f$ in figure~\ref{fig:weak}(a). At $a=0.50$ the evolved value is $0.05797\pm0.00282$, compared with $0.05502$ from the leading term. At $a=1$ the leading term falls short of the measured contrast by 2.27 percentage points (approximate 95 percent interval $[1.32,3.22]$ points, combining sampling and integration errors). The signed joint volume contrast departs from its leading prediction in the opposite direction (figure~\ref{fig:weak}, panel b). The leading term predicts $0.05980$ at $a=1$, whereas the maps give $0.04809\pm0.00145$. The change of sign in the departure illustrates the effect of fixing the marginal abundance.

The resolution test in table~\ref{tab:weakresolution} uses the same eight initial pairs at every resolution. At $a=0.25$ the main rank measurement lies 0.37 points below the leading prediction. The shift from $48^3$ to $72^3$ is already 0.39 points, so the early difference does not resolve a physical correction of that size. At $a=1$ the shift from $72^3$ to $96^3$ is 0.07 points in $C_f$. Moving the start from $1/16$ to $1/32$ at $72^3$ while retaining the logarithmic step size changes the final $C_f$ by $-0.09\pm0.02$ points. These shifts are smaller than the final excess over the leading prediction.

\begin{table}[tp]
\centering\small
\caption{Changes in the two observables under spatial refinement at identical comoving radii, using eight shared seed pairs. Entries are finer minus coarser in percentage points with one standard error on the paired change. Rank thresholds are recomputed at each resolution and in each jackknife sample.}
\label{tab:weakresolution}
\begin{tabular}{@{}crrrr@{}}\toprule
& \multicolumn{2}{c}{$100\,\Delta C_f$} & \multicolumn{2}{c}{$100\,\Delta A_0$}\\
\cmidrule(lr){2-3}\cmidrule(l){4-5}
$a$ & $48^3$ to $72^3$ & $72^3$ to $96^3$ & $48^3$ to $72^3$ & $72^3$ to $96^3$\\ \midrule
0.25 & $0.389\pm0.159$ & $0.065\pm0.019$ & $0.281\pm0.012$ & $0.073\pm0.006$\\
0.50 & $0.324\pm0.140$ & $0.129\pm0.035$ & $0.124\pm0.009$ & $0.068\pm0.004$\\
1.00 & $0.209\pm0.067$ & $0.071\pm0.034$ & $0.073\pm0.007$ & $0.052\pm0.005$\\ \bottomrule
\end{tabular}
\end{table}

The leading gravitational bispectrum therefore predicts the sign and weak amplitude scale of the rank asymmetry for this spectrum. Once the linear rms at the inner radius reaches $0.5$, the measured rank contrast exceeds that prediction while the signed joint volume contrast falls below its own prediction. The vanishing first variation of total disagreement makes its leading term a second order calculation.

\section{The first even correction to class disagreement}
\label{sec:even}

Section~\ref{sec:weak} showed that sign reversal forbids a correction linear in the growth amplitude to total class disagreement at zero threshold. We calculate the leading even coefficient with a Gaussian reference at the ensemble covariance. At weak amplitude we test it with the same reference. We follow a separate, per box residual through the final epoch.

\subsection{Cumulant order after matching the covariance}

In the whitened variables of section~\ref{sec:weak}, write the growing tensor expansion as $Z(a)=Z+aY+a^2U+\cdots$. Here $Z,Y,U$ are respectively linear, quadratic and cubic in the Gaussian initial modes. They have zero mean because the spatially constant density mode is removed. The covariance of $Z(a)$ is $\mathbb I+a^2V+\cdots$, with
\begin{equation}
 V_{ij}=\EE_0[Y_iY_j+Z_iU_j+U_iZ_j].
 \label{eq:evencov}
\end{equation}
The cubic cumulant is $aK+O(a^3)$ with $K$ as in eq.~\eqref{eq:treeK}. Write $a^2J^{(4)}$ for the leading connected fourth cumulant. Its coefficient consists of the six permutations of the slot choices in $\langle YYZZ\rangle_{\rm c}$ and the four in $\langle UZZZ\rangle_{\rm c}$. The covariance terms in these connected moments use the expansions in eq.~\eqref{eq:evencov}. Cumulants of order five and higher require more than two powers of the growth amplitude. In a Gaussian Wick expansion, connecting $n$ centred insertions needs at least $n-1$ pairings. The total polynomial degree at order $a^j$ is $n+j$, so a connected term requires $j\ge n-2$.

Let $\phi$ be the standard Gaussian density and define the Hermite tensors by
\begin{equation}
 H_{i_1\ldots i_r}=(-1)^r\phi^{-1}\partial_{i_1}\cdots\partial_{i_r}\phi.
\label{eq:hermite_definition}\end{equation}
Expanding the logarithm of the characteristic function near Gaussianity and returning to the density gives the order $a^2$ weight
\begin{equation}
 \frac12V_{ij}H_{ij}
   +\frac{1}{24}J^{(4)}_{ijkl}H_{ijkl}
   +\frac{1}{72}K_{ijk}K_{lmn}H_{ijklmn},
 \label{eq:even_all}
\end{equation}
as an asymptotic expansion about the Gaussian law \cite{sellentin}. The Gaussian field with the same ensemble covariance contributes exactly the first term. If $E$ is the event of a class change, then the residual for that ensemble Gaussian reference has the leading expression
\begin{equation}
 a^2\,\EE_0\!\left[
 \mathbf1_E\left(\frac{J^{(4)}_{ijkl}}{24}H_{ijkl}
        +\frac{K_{ijk}K_{lmn}}{72}H_{ijklmn}\right)\right]+o(a^2).
 \label{eq:even_res}
\end{equation}
Both terms are even under simultaneous sign reversal. The quadratic density gives the leading $K$, whereas the connected fourth cumulant also receives a contribution from the cubic density through $\langle UZZZ\rangle_{\rm c}$. Equation~\eqref{eq:even_res} compares the map with one Gaussian reference at the ensemble covariance. With a separate Gaussian reference for each box, the nonlinear dependence on covariance produces an additional term from fluctuations of its first derivative (appendix~\ref{app:density}). We test each residual with its corresponding covariance prescription.

\subsection{Perturbative maps and the covariance control}

We calculate the second and third order growing density from the EdS equations \cite{bern}. For two discrete fields in Fourier space write $(u\star_\gamma v)(\boldsymbol k)=\sum_{\boldsymbol p+\boldsymbol q=\boldsymbol k}\gamma(\boldsymbol p,\boldsymbol q)u(\boldsymbol p)v(\boldsymbol q)$. Define
\begin{equation}
 \alpha(\boldsymbol p,\boldsymbol q)=\frac{(\boldsymbol p+\boldsymbol q)\cdot\boldsymbol p}{p^2},
 \qquad
 \beta(\boldsymbol p,\boldsymbol q)=\frac{|\boldsymbol p+\boldsymbol q|^2\,\boldsymbol p\cdot\boldsymbol q}{2p^2q^2}.
 \label{eq:ptkernels}
\end{equation}
For $\theta=-\nabla\cdot\boldsymbol v/(aH)$, with $\boldsymbol v$ the physical peculiar velocity and $\nabla$ comoving, the growing coefficients obey
\begin{align}
 A_j&=\sum_{m=1}^{j-1}\theta_m\star_\alpha\delta_{j-m},&
 B_j&=\sum_{m=1}^{j-1}\theta_m\star_\beta\theta_{j-m},\notag\\
 \delta_j&=\frac{(2j+1)A_j+2B_j}{(2j+3)(j-1)},&
 \theta_j&=\frac{3A_j+2jB_j}{(2j+3)(j-1)} .
 \label{eq:ptrecurrence}
\end{align}
Here $\delta_1=\theta_1$ and constant modes are zero. The quadratic maps use $\delta_1+a\delta_2$. Cubic maps also include $a^2\delta_3$. They represent $\delta(a)/a$. We apply the tidal projection and smoothing to the summed field. The input cutoff confines all products through third order to modes with magnitude at most 24 times the box fundamental, and the Nyquist planes are discarded at mesh size 48. We checked eq.~\eqref{eq:ptrecurrence} against direct Fourier convolution with nonparallel modes and against the exact planar growing solution before shell crossing.

For every map $X$ let $R(X)=D(X)-D_{\rm G}(\widehat\Sigma_X)$, where $\widehat\Sigma_X$ is the full realised covariance from eq.~\eqref{eq:matchedcov}. We compare
\begin{equation}
 \mathcal R(a)=\EE_{\rm seed}\left[
   \tfrac12\sum_{s=\pm1}\big\{R(X_s(a))-R(X_s(0))\big\}\right].
 \label{eq:even_measure}
\end{equation}
The subtraction uses the same linear phases and removes the residual at zero amplitude for each box. For the truncated fields $X_-(a)=-X_+(-a)$, so the sign average in eq.~\eqref{eq:even_measure} is even in the signed amplitude before averaging over seeds. Dividing the measured change by $a^2$ tests the approach to weak amplitude scaling.

For the reference probability we integrate the common identity displacement as in section~\ref{sec:weak}. For each draw in the other eleven coordinates, the three eigenvalues at each radius define four intervals for the Gaussian displacement. The sum of the Gaussian probabilities of intersections of intervals with the same class is the conditional agreement probability. We use 4096 scrambled Sobol points per seed and share them across fields and signs. A higher sampling check uses two independent scrambles with 32,768 points each. We also use two phase maps with exact amplitude preservation at the three epochs shared with the particle calculation.

\subsection{Direct evaluation of the even coefficient}

\begin{figure}[tp]
\centering\includegraphics[width=.99\linewidth]{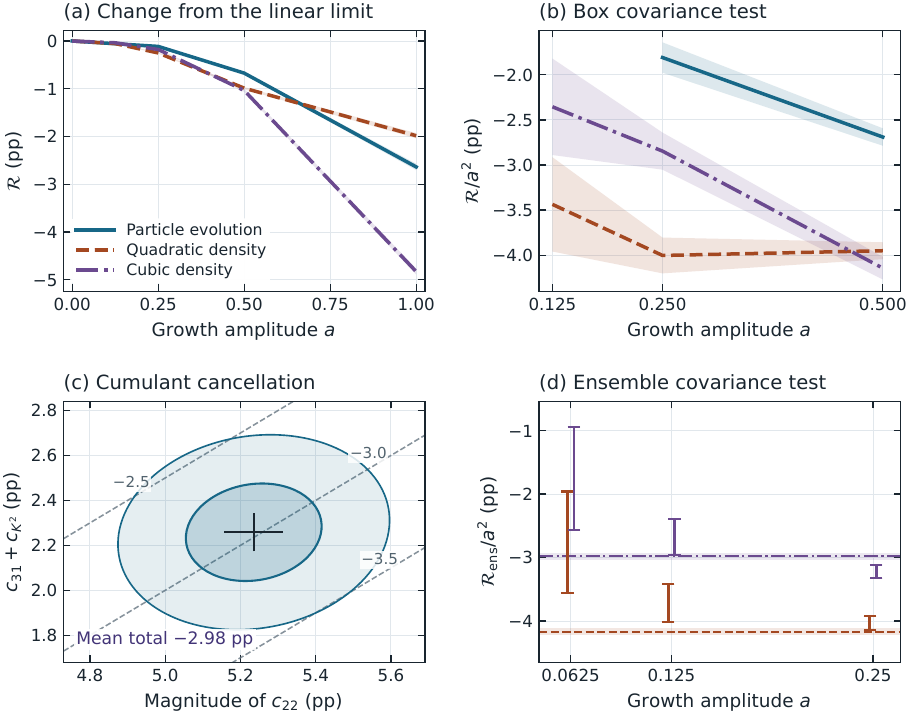}
\caption{Even response of class disagreement. (a) Change in the residual from its linear baseline and (b) change divided by $a^2$, using the Gaussian reference for each box. Bands give one standard error across 48 seed pairs. (c) Paired estimates of the magnitude $-c_{22}$ and the compensation $c_{31}+c_{K^2}$ from the 24 integration scrambles at ensemble covariance. The crosshair marks their mean. Ellipses have radii one and two in the metric of the sample covariance and describe variation across scrambles. Dashed lines mark constant total $c_{22}+c_{31}+c_{K^2}$. (d) Scaled residuals from 128 new seed pairs using the ensemble reference. Horizontal lines and shading show the predicted coefficients and their integration errors. Line colours and styles in (d) follow (a). Finite amplitude error bars show one standard error and are offset horizontally for clarity. The axes labelled pp are in percentage points.}
\label{fig:even_disagreement}
\end{figure}

For the reference spectrum with $n=-1$ and radii $L/24$ and $L/16$, we compute the contractions in eq.~\eqref{eq:even_res} by Wick expansion on the periodic Fourier lattice. Appendix~\ref{app:wick} gives the contractions and Gaussian integration. The three contributions to the coefficient of $a^2$, in probability units, are $-0.05235\pm0.00037$ from $\langle YYZZ\rangle_{\rm c}$, $+0.01193\pm0.00005$ from $\langle UZZZ\rangle_{\rm c}$ and $+0.01066\pm0.00043$ from $K^2$. Their sum is $-0.02977\pm0.00054$, where the error is the standard error of the sum across 24 independent scrambles. Let $c_{22}$, $c_{31}$ and $c_{K^2}$ denote these three contributions in percentage points. The joint variation across scrambles of the magnitude of $c_{22}$ and the positive compensation $c_{31}+c_{K^2}$ is shown in figure~\ref{fig:even_disagreement}(c). The net contribution is their signed sum. A separate twelve dimensional integral without conditioning the common Gaussian displacement gives $-0.03054\pm0.00147$. Three projections of the mixed cumulants also agree with direct real space estimates from 128 independent Gaussian fields.

To test the prediction at finite amplitude with the same reference, we generate 128 new inverted pairs. We measure the changes in the map disagreement and in each box covariance relative to its linear member. We add the measured mean change to the exact linear covariance before evaluating the Gaussian probability. Denoting the resulting residual by $\mathcal R_{\rm ens}$, the cubic density gives $\mathcal R_{\rm ens}/a^2=-0.02676\pm0.00281$ at $a=0.125$ and $-0.03217\pm0.00106$ at $a=0.25$. At $a=0.0625$ the estimate is $-0.01752\pm0.00812$. The measurements are compared with the leading coefficients for both density truncations in figure~\ref{fig:even_disagreement}(d). At $a=0.125$, the measurement and the cubic prediction differ by about one combined standard error. At $a=0.25$ the central estimate lies below the leading prediction. The errors combine the independent seed uncertainty, including covariance estimation, and the Gaussian integration error.

\subsection{Evolution with a reference for each box}

The main ensemble has 48 fresh inverted seed pairs with $n=-1$, $R_1=L/24$ and $R_2=L/16$. We report the changes relative to the linear residual in table~\ref{tab:even_amplitudes} and figure~\ref{fig:even_disagreement}(a). In the cubic field it is negative at both weak amplitudes. At $a=0.125$ and $a=0.25$, the values of $\mathcal R(a)/a^2$ are $-0.0236\pm0.0053$ and $-0.0285\pm0.0021$, respectively [figure~\ref{fig:even_disagreement}(b)]. Their paired difference is $-0.0049\pm0.0041$. The estimates are consistent with a common leading coefficient at this precision. At $a=0.25$, including the cubic density shifts the residual by $+0.072\pm0.005$ percentage points, reducing the magnitude of the deficit. The finer Gaussian integration changes the weak residuals by less than $0.0003$ points in the mean.

\begin{table}[tp]
\centering\small
\caption{Change in the disagreement residual relative to the corresponding linear map, in percentage points. Entries give means and one standard error from 48 inverted seed pairs. The cubic density includes both nonlinear terms of eq.~\eqref{eq:ptrecurrence}. Particle outputs are recorded at the three indicated epochs.}
\label{tab:even_amplitudes}
\begin{tabular}{@{}crrr@{}}\toprule
$a$ & Quadratic density & Cubic density & Particle evolution\\ \midrule
0.125 & $-0.054\pm0.008$ & $-0.037\pm0.008$ & \multicolumn{1}{c}{\textemdash}\\
0.25  & $-0.250\pm0.012$ & $-0.178\pm0.013$ & $-0.113\pm0.011$\\
0.50  & $-0.987\pm0.023$ & $-1.036\pm0.031$ & $-0.673\pm0.024$\\
1.00  & $-1.983\pm0.055$ & $-4.829\pm0.063$ & $-2.636\pm0.067$\\ \bottomrule
\end{tabular}
\end{table}

Spatial refinement has its largest relative effect on the early particle result. For eight shared seeds at $a=0.25$ the change from $48^3$ to $72^3$ is $-0.051\pm0.037$ points and the change from $72^3$ to $96^3$ is $0.001\pm0.015$ points. At $a=1$ the corresponding shifts are $-0.082\pm0.035$ and $-0.021\pm0.009$ points. Moving the start of the $72^3$ run from $1/16$ to $1/32$ at fixed logarithmic step size shifts the final result by $+0.011\pm0.003$ points. These tests use the same input modes and smoothing radii.

At $a=1$ the Gaussian reference for the evolved field changes by $+2.55\pm0.05$ percentage points while the evolved maps change by $-0.08\pm0.05$ points, relative to the respective linear baselines. The residual change against phase randomisation is $-2.65\pm0.08$ points. In the cubic field, the larger disagreement of its Gaussian reference is accompanied by a rise of $2.05\pm0.11$ points in the map itself. In the main ensemble the cubic residual is $2.19\pm0.05$ points more negative than the evolved residual (95 percent paired interval for the signed difference $[-2.30,-2.08]$). The excess persists for the common seed subset when the particle calculation is refined.

For these filters and this input spectrum, cubic order in the density matters even in the weak response. At larger amplitude both the relative effect of the cubic term and the evolution of the Gaussian reference change. The particle residual has the same sign as the weak response. Extrapolating the cubic density to the final epoch overestimates the deficit and changes the smoothed power.

\section{Sensitivity to the expansion history at fixed linear growth}
\label{sec:history}

To test whether the disagreement deficit and the difference in extreme environment retention survive a change of background, we evolve identical initial modes in two flat cosmologies. The expansion laws have present matter fractions $\Omega_{\rm m0}=1$ and $0.315$, respectively. Both histories use collisionless matter and share the shape, amplitude and phases of the linear density extrapolated to the final epoch. The different expansion laws therefore isolate the effect of the background for this common linear input.

\subsection{Common initial spectrum and growth variable}

We adopt a smooth cold dark matter spectrum $P_{\rm L}(k)=A k^{n_{\rm s}}T_0^2(k)$, with the zero baryon transfer function of Eisenstein and Hu \cite{eh}. In terms of $q=k/(13.41\,k_{\rm eq})$ this function is
\begin{equation}
 T_0(q)=\frac{L(q)}{L(q)+C(q)q^2},\qquad
 L(q)=\ln(2e+1.8q),\qquad C(q)=14.2+\frac{731}{1+62.5q}.
\label{eq:transfer}\end{equation}
Here $k_{\rm eq}=0.0746\,\omega_{\rm m}(T_{\rm CMB}/2.7\,{\rm K})^{-2}\,{\rm Mpc}^{-1}$ and both wave numbers in $q$ are in the same units. The shape parameters are $n_{\rm s}=0.965$, $h=0.674$ and $\omega_{\rm m}=0.315h^2$, with a microwave background temperature of $2.7255\,{\rm K}$. The spectral index, matter density and amplitude are chosen near the Planck values \cite{planck}. The same smooth spectrum sets the initial conditions in both histories. We set $A$ by requiring $\sigma_8=0.811$ for the continuum spectrum with a spherical top hat window of radius $8\,h^{-1}{\rm Mpc}$. The Fourier initialisation retains $0<k\leq0.524\,h\,{\rm Mpc}^{-1}$ in both backgrounds. The upper cutoff is 16 times the fundamental of the reference cube of side $192\,h^{-1}{\rm Mpc}$. Changing the size of the cube changes the set of sampled wavelengths, while the input normalisation and comoving cutoff are held fixed. The two Gaussian smoothing radii are $8$ and $12\,h^{-1}{\rm Mpc}$. The expected linear density rms at the smaller radius on the reference lattice is 0.435.

Let $E(a)=H(a)/H_0$ and $\Omega_{\rm m}(a)=\Omega_{\rm m0}a^{-3}/E^2(a)$, where $E^2=\Omega_{\rm m0}a^{-3}+1-\Omega_{\rm m0}$. In this section $\boldsymbol p$ denotes the canonical momentum of eq.~\eqref{eq:pm} divided by $H_0$. The dynamics are
\begin{equation}
 \frac{d\boldsymbol x}{da}=\frac{\boldsymbol p}{a^3 E(a)},
 \qquad
 \frac{d\boldsymbol p}{da}=\frac{3\Omega_{\rm m0}}{2a^2 E(a)}\,\mathcal G[\delta](\boldsymbol x).
 \label{eq:background_dynamics}
\end{equation}
The operator $\mathcal G$ is as defined in eq.~\eqref{eq:pm}. We normalise the growing solution to $D_+(1)=1$ and use $\eta=\ln D_+$ as a common evolution coordinate. With $f_g=d\ln D_+/d\ln a$, eq.~\eqref{eq:background_dynamics} is equivalent to
\begin{equation}
 \frac{d^2\boldsymbol x}{d\eta^2}
 +\left(\frac32\kappa-1\right)\frac{d\boldsymbol x}{d\eta}
 =\frac32\kappa\,\mathcal G[\delta],\qquad
 \kappa=\frac{\Omega_{\rm m}(a)}{f_g^2}.
 \label{eq:background_clock}
\end{equation}
The linear displacement grows as $D_+$ in either background. An explicit dependence on the expansion history remains through $\kappa$ once the density and force become nonlinear, consistent with the time dependence of the nonlinear growing kernels \cite{fasiello}. In the matter plus $\Lambda$ history, $\kappa$ rises from 1.0015 at $D_+=0.25$ to 1.134 at $D_+=1$. In Einstein--de Sitter it is unity.

At initialisation we use the first and second order growing displacements. The linear growth factor obeys $D_+''+bD_+'-\tfrac32\Omega_{\rm m}D_+=0$, where primes denote $d/d\ln a$ and $b=2-\tfrac32\Omega_{\rm m}$. For the spatial quadratic displacement in eq.~\eqref{eq:ic} with the factor $3/7$ removed, the time coefficient satisfies
\begin{equation}
 F''+bF'-\frac32\Omega_{\rm m}F=\frac32\Omega_{\rm m}D_+^2,\qquad
 \lim_{a\to0}\frac{F}{D_+^2}=\frac37 .
 \label{eq:background_2lpt}
\end{equation}
The initial momentum follows by differentiating both growing amplitudes. We form the quadratic source on a Fourier grid that resolves its doubled bandwidth before restricting the displacement to the retained mesh modes. At the final epoch $F/D_+^2$ is 0.4321 in the matter plus $\Lambda$ history and $3/7$ in Einstein--de Sitter. We tested the linear solution against the independent growth integral \cite{heath} and the particle integration against the planar growing displacement in both backgrounds.

\subsection{Paired measurements}

For each of 32 independent initial seeds and its inverted partner, we evolve $48^3$ particles on a $48^3$ force mesh in the reference cube. Both backgrounds start at $D_+=1/16$ and use 24 equal steps in $\ln D_+$. Each kick and drift integrates the corresponding coefficient in eq.~\eqref{eq:background_dynamics}. We measure at $D_+=0.25$, $0.50$ and $1$. In the $\Lambda$ history the first two outputs occur at $a=0.198$ and $0.404$. The measured tidal labels use zero threshold. We compute $C_f$ from the equal abundance rank masks of eq.~\eqref{eq:rank} with $f=0.079676$ separately in each box. Disagreement is compared with a Gaussian reference at the full measured covariance, using the common displacement integral of section~\ref{sec:even}, and with two independent phase controls that retain the measured Fourier amplitudes. Each paired difference keeps both signs and both backgrounds of an initial seed together.

At $D_+=1$, the reference cube gives $\Delta D=-1.599\pm0.095$ percentage points in Einstein--de Sitter and $-1.620\pm0.095$ in the $\Lambda$ history. The corresponding equal abundance retention contrasts are $10.248\pm0.358$ and $10.327\pm0.359$ points. Errors here are standard errors across seeds. Using the same initial phases reduces the sampling error of the difference below that of either mean. The disagreement residual changes by $-0.0210\pm0.0014$ points, with 95 percent interval $[-0.0238,-0.0181]$. The retention contrast changes by $+0.0789\pm0.0067$ points, with interval $[0.0652,0.0926]$. The difference in disagreement residual using the phase controls is $-0.0217\pm0.0016$ points.

\begin{figure}[tp]
\centering\includegraphics[width=.99\linewidth]{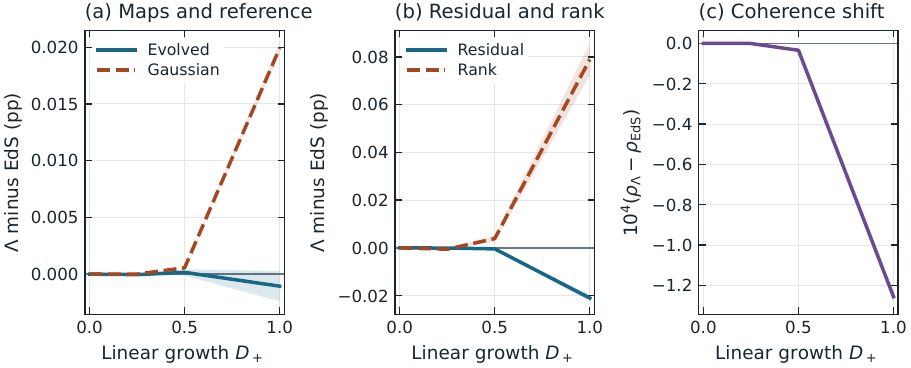}
\caption{Expansion history contrasts at fixed linear growth. All changes are $\Lambda$ minus Einstein--de Sitter, for identical initial phases and terminal linear power. (a) Differences in the evolved map disagreement and in its Gaussian reference. (b) Differences in the disagreement residual and in the equal abundance retention contrast. (c) Change in density coherence in units of $10^{-4}$. Lines join the common linear limit and three measured epochs. Bands give one standard error across 32 seed pairs.}
\label{fig:history}
\end{figure}

The changes in the evolved maps and their Gaussian references are shown separately in figure~\ref{fig:history}(a). At $D_+=1$ the absolute map disagreement changes by $-0.0011\pm0.0013$ points. The Gaussian reference increases by $0.01992\pm0.00041$ points. The measured density coherence decreases by $(1.26\pm0.02)\times10^{-4}$ in the $\Lambda$ history (figure~\ref{fig:history}(c)). An isotropic Gaussian calculation using each member's measured coherence gives a change of $0.01995\pm0.00041$ points. Its difference from the full lattice result is $0.00003\pm0.00005$ points. The shift in scalar coherence therefore accounts for the measured change in Gaussian disagreement to the stated precision. Near the final epoch, the $\Lambda$ history has a smaller disagreement residual and a larger retention advantage for voids (figure~\ref{fig:history}, panel b).

We test the history differences with a finer particle and force mesh, an earlier start at the same spacing in $\ln D_+$, and a cube with twice the side at fixed comoving particle spacing (table~\ref{tab:history}). The disagreement contrast retains its scale and sign in all checks. The earlier start reduces the retention difference by $0.0333\pm0.0128$ points on the same eight seeds. The shift exceeds the sampling error of the main estimate. The contrast remains positive with the earlier start and in the larger volume. These results concern the specified smoothing windows, amplitude and smooth input spectrum. For either background, both signals are much larger than their shifts between the two expansion histories.

\begin{table}[tp]
\centering\small
\caption{Paired history differences at $D_+=1$ in percentage points, with one standard error. Each difference is the value in the $\Lambda$ background minus that in Einstein--de Sitter. The finer mesh and earlier start use the first eight main seeds. The larger cube uses eight independent paired seeds.}
\label{tab:history}
\begin{tabular}{@{}lrrrr@{}}\toprule
Case & Particles & Seed pairs & $100\,(\Delta D_\Lambda-\Delta D_{\rm EdS})$ & $100\,(C_{f,\Lambda}-C_{f,\rm EdS})$\\ \midrule
Reference & $48^3$ & 32 & $-0.0210\pm0.0014$ & $+0.0789\pm0.0067$\\
Finer mesh & $72^3$ & 8 & $-0.0200\pm0.0024$ & $+0.0700\pm0.0081$\\
Earlier start & $48^3$ & 8 & $-0.0211\pm0.0019$ & $+0.0603\pm0.0092$\\
Double side & $96^3$ & 8 & $-0.0197\pm0.0012$ & $+0.0735\pm0.0044$\\ \bottomrule
\end{tabular}
\end{table}

\section{An independent test with Sahyadri}
\label{sec:external}

We measure both statistics in the fiducial volume of the public Sahyadri suite \cite{sahyadri,sahyadri_data}. It evolves $2048^3$ particles in a periodic cube of side $200\,h^{-1}{\rm Mpc}$ with independent phases, a different transfer function and its own force solver. The snapshot headers give $\Omega_{\rm m}=0.31377$ and $h=0.6736$. Each of the two public snapshots contains $85\,899\,346$ particle positions, one percent of the full realisation. We use snapshots 100 and 046, at $z=0$ and $z=0.99667$, which sample the same volume at two epochs. The primary smoothing radii are the same as in section~\ref{sec:history}.

We decode the particle positions using the format of the release and deposit them with a cloud in cell window on a $128^3$ periodic lattice. We remove the constant mode, deconvolve the assignment window and retain wavenumbers up to $0.8\,h\,{\rm Mpc}^{-1}$. The primary Gaussian radii are $8$ and $12\,h^{-1}{\rm Mpc}$ in comoving units. We apply the same zero threshold classification and the same rank fraction $f=0.079676$ as before. For $D_{\rm G}$ we use the full measured Fourier covariance in eq.~\eqref{eq:matchedcov}, including its directional dependence. The quoted integration errors at the primary radii come from 24 independent scrambled Sobol integrations with $2^{17}$ points each. We also generate 128 phase randomised fields at each epoch with the full realised mode amplitudes held fixed.

The primary measurements are listed in table~\ref{tab:external}. At $z=0$, $22.140$ percent of sites change class, a deficit of $1.091$ percentage points relative to the Gaussian reference. The deficit relative to the mean of the fixed amplitude phase controls is $1.107\pm0.046$ points, where the error is the standard error over the control draws. Extreme voids retain $3.039$ percentage points more of their volume than knots at equal abundance. This is smaller than the $10.327$ points measured with the same radii in the controlled $\Lambda$ ensemble of section~\ref{sec:history}.

\begin{table}[tp]
\centering\small
\caption{Sahyadri at Gaussian radii of $8$ and $12\,h^{-1}{\rm Mpc}$, with $128^3$ analysis cells. The first two probability columns are in percent. The last two columns are in percentage points. Errors on $D_{\rm G}$ and $\Delta D$ describe the Gaussian integration. Both snapshots belong to one periodic realisation.}
\label{tab:external}
\begin{tabular}{@{}crrrr@{}}\toprule
$z$ & $100D_{\rm map}$ & $100D_{\rm G}$ & $100\Delta D$ & $100C_f$\\ \midrule
0.997 & 22.973 & $22.944\pm0.003$ & $+0.029\pm0.003$ & $-0.866$\\
0 & 22.140 & $23.231\pm0.003$ & $-1.091\pm0.003$ & $+3.039$\\ \bottomrule
\end{tabular}
\end{table}

\begin{figure}[tp]
\centering\includegraphics[width=.99\linewidth]{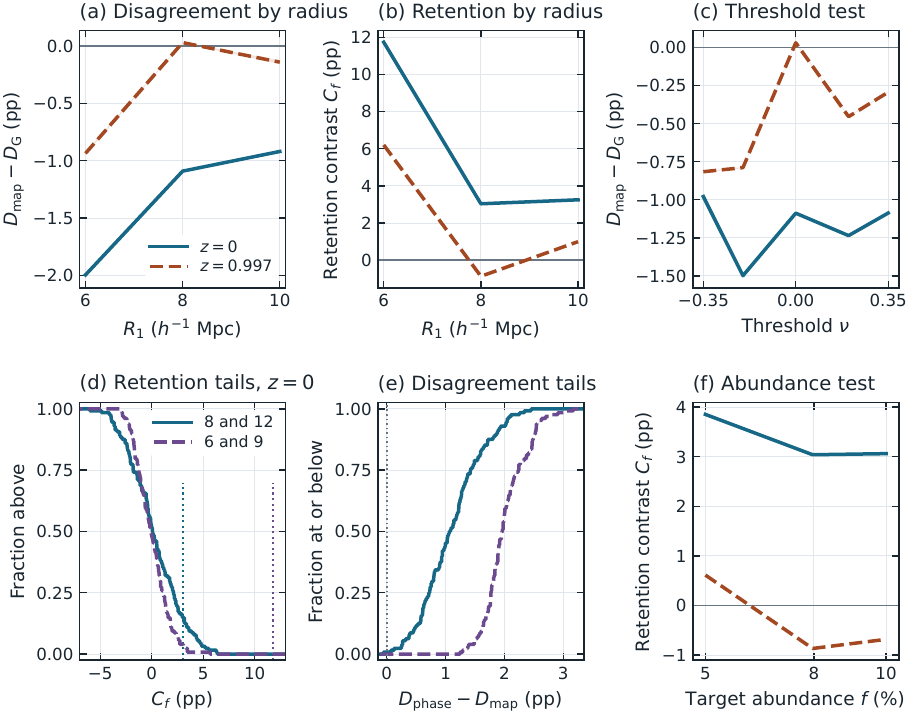}
\caption{Sahyadri measurements at two epochs. (a) Disagreement residual and (b) retention contrast versus $R_1$, with $R_2=1.5 R_1$. (c) Threshold response at $8$ and $12\,h^{-1}{\rm Mpc}$. Bands in (a) and (c) give one integration standard error. The zero threshold estimates at the primary radii in (a) and (c) come from independent scrambles. Panels (d) and (e) compare 128 phase controls per radius pair at $z=0$. In (d) the vertical dotted lines mark the measured retention contrasts. Panel (e) gives the cumulative distribution of $D_{\rm phase}-D_{\rm map}$. Zero marks the measured disagreement for each pair. The legend in (d) applies to both panels, with radii in $h^{-1}{\rm Mpc}$. (f) Retention contrast versus abundance at the primary radii. Redshift styles follow (a) in panels (b), (c) and (f).}
\label{fig:external}
\end{figure}

At $z=0$, one of the 128 phase controls gives a disagreement at or below the measured value, and 19 give a retention contrast at or above the measured value. Including the evolved field in the count gives one sided Monte Carlo ranks of $2/129$ and $20/129$, respectively. At $z=0.997$ the corresponding ranks are $66/129$ and $82/129$. These ranks characterise the conditional phase ensemble for the measured power of this particular volume. The two empirical phase distributions are shown in figure~\ref{fig:external}, panels (d) and (e).

We test two other radius pairs at the same ratio $R_2/R_1=1.5$ (figure~\ref{fig:external}, panels a and b). At $z=0$, the smaller pair of $6$ and $9\,h^{-1}{\rm Mpc}$ gives $\Delta D=-1.994$ and $C_f=11.741$ percentage points. At this smaller pair, none of the 128 phase controls at either epoch reaches the observed low disagreement or high retention contrast. The smaller pair was chosen for an exploratory comparison after the primary test. It gives larger contrasts within the same realisation. Keeping the primary radii at $z=0$, increasing the analysis mesh from $128^3$ to $256^3$ changes $\Delta D$ by $-0.0054$ points and $C_f$ by $-0.060$ points. Shifting and interlacing the assignment modifies the two values by less than $0.002$ points each. Using two disjoint random halves of the released particles changes each primary estimate by at most $0.026$ points. These tests assess numerical sensitivity within the fiducial volume.

We also vary the normalised eigenvalue threshold at the primary radii, using an unnormalised eigenvalue cut of $\nu\sigma_s$ at radius $s$. The map and its Gaussian control use the same normalisation (figure~\ref{fig:external}, panel c). At $z=0$, $\Delta D$ is negative at all five thresholds and ranges from $-1.498$ to $-0.979$ percentage points. Near redshift one it is slightly positive at zero threshold and negative at the four nonzero thresholds tested. At the primary radii the retention contrast at $z=0$ ranges from $3.039$ to $3.852$ points for rank fractions of $0.05$, $0.079676$ and $0.10$. Near redshift one the rank contrast changes sign over those fractions (figure~\ref{fig:external}(f)).

\FloatBarrier

\section{Discussion}
\label{sec:discussion}

The Gaussian calculation separates two aspects of scale dependence. A normalised threshold fixes the class fractions at each radius, while the correlation between the radii controls whether the same sites retain their class. The form of the smoothing window determines how quickly that correlation changes as the radii separate. In the evolved ensembles, the Gaussian and phase controls give similar disagreement deficits. Matching the density rms and coherence across the spectral tilts leaves a measurable difference in that deficit.

At zero threshold, sign symmetry removes the term linear in the third cumulant from the total disagreement but allows it in the retention contrast. For the reference spectrum, the squared third cumulant partly offsets the fourth cumulant contribution to disagreement at order $a^2$. The leading coefficient applies to a Gaussian reference at the ensemble covariance. The evolved particle measurements instead use each box's covariance, and their late time departure from cubic density perturbation theory limits the extrapolation of the weak amplitude prediction.

In Sahyadri, the disagreement deficit appears at redshift zero at the primary radii, while the retention contrast is larger at the smaller, exploratory pair. Both snapshots sample the same volume. The phase ranks condition on its realised spectrum and do not measure scatter between volumes. The controlled history comparison varies expansion at fixed linear power, whereas Sahyadri also changes the spectrum and volume. The statistics quantify agreement and retention of the volume weighted matter field. Environment dependent power spectra have been used to extract cosmological information from simulations \cite{bonnaire}. Applying the present statistics to a tracer catalogue would require a density reconstruction and treatment of the selection effects.

\section{Conclusions}
For Gaussian tidal tensors, the eigenvalue densities at the threshold give both the leading disagreement at nearby smoothing radii and a bound at finite angular separation. With gravitational evolution at zero threshold, the disagreement corrected for the ensemble covariance responds at order $a^2$ for the reference spectrum. The retention contrast at fixed abundance responds at order $a$. In one independent Sahyadri realisation at redshift zero, the disagreement deficit persists over the five thresholds tested at the primary radii, while the retention contrast varies with the smoothing scale.

\paragraph{Data Availability Statement.}
The electronic supplement contains transition matrices, seed level summaries, example Fourier fields and figure data. The Sahyadri particle positions and snapshot metadata are available from the public release \cite{sahyadri_data}. The electronic supplement provides filtered Fourier inputs and checksums for the public files used in section~\ref{sec:external}.

\paragraph{Code Availability Statement.}
The electronic supplement contains the simulation, analysis and figure scripts, together with instructions for reproducing the reported results. It also specifies the versions of the numerical and plotting dependencies \cite{numpy,scipy,pandas,matplotlib}.

\appendix
\section{Expansion of class probabilities and covariance references}
\label{app:density}

On a fixed Fourier lattice with nonsingular linear tidal covariance, decompose the standardised Gaussian input into the linear tidal coordinates $z$ and their independent orthogonal complement $w$. To second order in the growth amplitude the tidal vector is $z+aY(z,w)+a^2U(z,w)$, with polynomial $Y$ and $U$. Choose a smooth cutoff of the perturbation in $(z,w)$ at radius $B(a)=c\sqrt{\log(1/|a|)}$ with $c$ large enough that omitted Gaussian tails are $o(a^2)$, including polynomial weights. For each $w$ the resulting map is the identity plus a smooth perturbation, with a Jacobian that differs from the identity by $O(|a|B^m)$ for some finite $m$. It is globally invertible for sufficiently small $|a|$. Expanding the inverse and its determinant in the change of density gives an integrated remainder bounded by $O(|a|^3 B^{m'})$ for some finite $m'$, plus the omitted Gaussian tails. Removing the cutoff in the coefficients gives an $L^1$ density expansion through order $a^2$ for the polynomial map. This justifies integration against bounded classification indicators on the finite lattice. The coefficients follow by expanding the characteristic function and give eqs.~\eqref{eq:score} and \eqref{eq:even_all}. Finite amplitude particle evolution is tested separately.

For the reference evaluated separately for each box, expand the box covariance as $\widehat\Sigma(a)=\widehat\Sigma_0+a\widehat\Sigma_1+a^2\widehat\Sigma_2+\cdots$. On positive definite covariances the Gaussian probability $g(\Sigma)=D_{\rm G}(\Sigma)$ is smooth by differentiation of its normal density. For a fixed box, the coefficient at order $a^2$ is
\begin{equation}
 g'_{\widehat\Sigma_0}[\widehat\Sigma_2]
 +\frac12\,g''_{\widehat\Sigma_0}[\widehat\Sigma_1,\widehat\Sigma_1].
 \label{eq:conditional_cov_response}
\end{equation}
Inverting the linear input changes the sign of $\widehat\Sigma_1$, so the linear term cancels within a pair while the Hessian term in eq.~\eqref{eq:conditional_cov_response} survives. In the ensemble reference the first covariance derivative vanishes before evaluating $g$. The two procedures differ even after subtraction of each linear baseline.

\section{Contractions for the even coefficient}
\label{app:wick}

Use the real even Fourier filters $B_i$ from section~\ref{sec:weak}. Let $P_{\boldsymbol q}=\EE|d_{1,\boldsymbol q}|^2$ be the variance of each linear Fourier series coefficient. Then $r_i(\boldsymbol q)=P_{\boldsymbol q}B_i(\boldsymbol q)$ is the shift of the input mode under a unit tilt by $Z_i$. Define
\begin{equation}
 T_{ij}(\boldsymbol p)
   =\sum_{\boldsymbol q} B_i(\boldsymbol p+\boldsymbol q)
      F_2(\boldsymbol p,\boldsymbol q)r_j(\boldsymbol q).
 \label{eq:fourth_t}
\end{equation}
The connected part for two specified quadratic insertions is
\begin{align}
 \operatorname{cum}(Y_i,Y_j,Z_k,Z_l)
   &=4\sum_{\boldsymbol p}P_{\boldsymbol p}
       \big[T_{ik}(\boldsymbol p)T_{jl}(-\boldsymbol p)
          +T_{il}(\boldsymbol p)T_{jk}(-\boldsymbol p)\big].
 \label{eq:fourth_yyzz}
\end{align}
A linear leg attaches to each quadratic vertex, and the remaining two legs connect the vertices. Disconnected pairings are removed in the cumulant. For a cubic insertion, let $U_i[d]$ be the homogeneous cubic functional of the input modes obtained from eq.~\eqref{eq:ptrecurrence}, followed by the filter $B_i$. The connected part is the coefficient of $uvw$ in $U_i[u r_j+v r_k+w r_l]$. It is evaluated by cubic polarisation, which retains the three connections to the linear insertions. Summing over the six placements of the quadratic pair and the four placements of the cubic vertex gives the corresponding components of $J^{(4)}$. The filters in both contractions use the full lattice covariance, including its anisotropy.

For the sixth order Hermite contraction in eq.~\eqref{eq:even_res}, set $h_3=K:H_3$, $A_{abcd}=\operatorname{Sym}\sum_i K_{iab}K_{icd}$ and $B_{ab}=\sum_{ij} K_{ija}K_{ijb}$. Here $\operatorname{Sym}$ averages the permutations of the four free indices. The Gaussian product identity gives
\begin{equation}
 (K\otimes K):H_6=h_3^2-9 A:H_4-18 B:H_2-6\|K\|^2 .
 \label{eq:hermite_product}
\end{equation}
We integrate the common identity displacement analytically as in section~\ref{sec:weak}. In each interval of that normal coordinate the event indicator is constant and the score is a polynomial of degree at most six. Its truncated moments follow by integrating the normal density by parts. We evaluate the remaining eleven dimensional expectation with 24 independent scrambles of $2^{16}$ points each. The independent direct integral draws all twelve coordinates with 24 scrambles of $2^{17}$ points.

For the finite amplitude uncertainty, covariance fluctuations enter through the derivative
\begin{equation}
 g'_\Sigma[H]=\frac12\,\EE_\Sigma\!\left[
 \mathbf1_E\bigl(x^{\mathsf T}\Sigma^{-1}H\Sigma^{-1}x-
                     \operatorname{tr}(\Sigma^{-1}H)\bigr)\right].
 \label{eq:gauss_influence}
\end{equation}
Contracting this derivative with each paired covariance change accounts to first order for the dependence of the ensemble Gaussian reference on the estimated covariance. The point estimate uses $g$ evaluated at the shifted covariance.

\end{document}